# Decision Support Systems Architectures

**Cristina Ofelia Stanciu**
**„Tibiscus" University of Timişoara, Romania**

REZUMAT. Lucrarea prezintă principalele componente ale sistemelor de asistare a deciziei, apoi sunt descrise, analizate respectiv comparate trei tipuri de arhitecturi ale acestor sisteme: arhitectura în reţea, arhitectura centralizată şi arhitectura ierarhizată.

According to Sprague and Carlson [Lu03], decision support systems would consist in the following components (Figure 1): data management component; model management component; user interface management component; decision support system architecture.

Nowadays, the components of decision support systems are very much like the ones identified by Sprague in 1982 (Figure 1): user interface; knowledge based subsystems; data management module; model management module.

The user interface is a component that provides the communication between the user and the decision support system. The proper design of this component is really important, as it is the only one the user actually deals with.

The data management method is a subsystem of the computer-based decision support system, and has a number of subcomponents of its own (Figure 2.):
- the integrated decision support system database, which includes data extracted from internal and external sources, data which can be maintained in the database or can be accessed only when is useful;
- the database management system; the database can be relational or multidimensional;

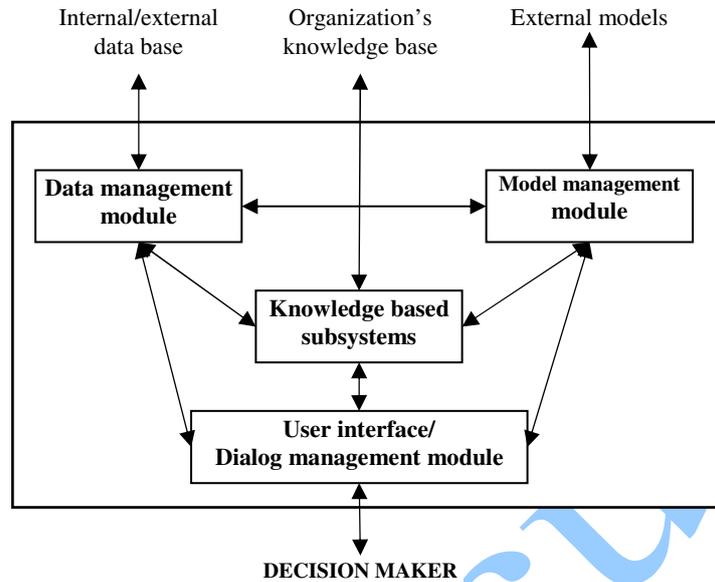

**Figure 1.** Decision support systems' components
Adapted after [Lu03])

- a data dictionary, implying a catalog containing all the definitions of database data; it is used in the decisional process identification and definition phase;
- query tools, assuming the existence of languages for querying databases.

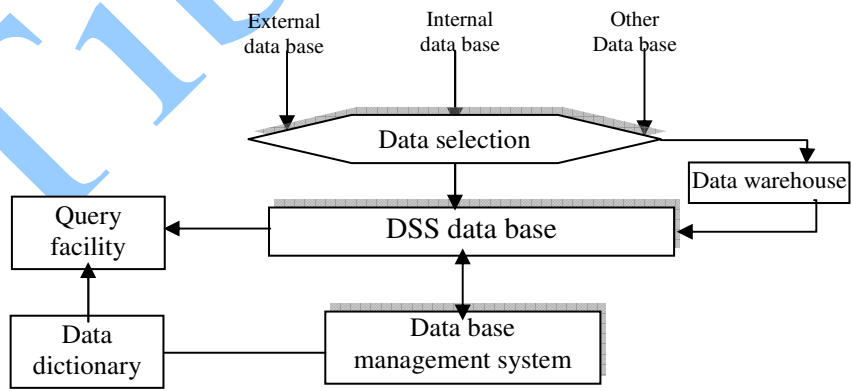

**Figure 2.** Data management module
(Adapted after [TA01])

The model management module consists in the following components (Figure 3):
- the model base, that contains the quantitative models that offer the system the capacity of analyzing and finding solutions to problems [ZAB01];
- the model base management module, that is meant to create new models by using programming languages;
- the model dictionary, that contains the models' definition and other information related tot hem;
- the creation, execution and integration module of models, that will interpret the user's instructions according to models and will transfer them towards the model management system [ZAB01].

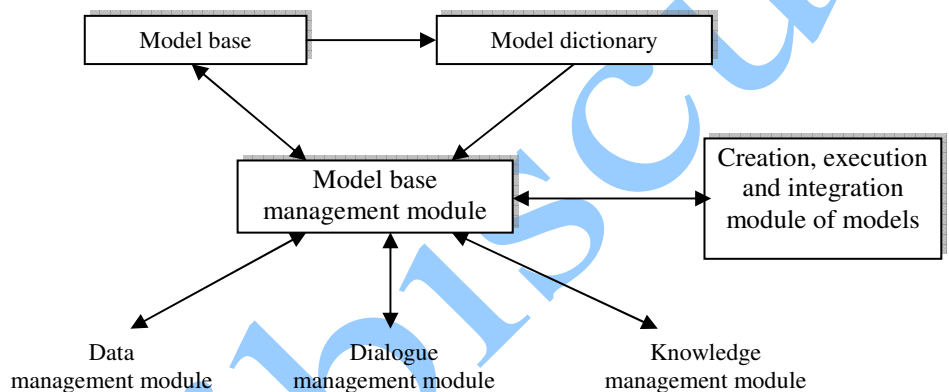

**Figure 3.** Model management module
(Adapted after [TA01])

**The network architecture**

In the network architecture (Figure 4.), each model has its own data base, its own integration model and its own dialogue module. This way, the assembly formed by the dialogue module, the model and the data base forms a complex, similar to a network station, and these complexes are controlled by an integration unit.

This architecture type has a high level of modularity and is at the same time an open and adaptable architecture, and its main advantage is that in case a complex suffers changes, these changes will not be transmitted to other complexes.

343

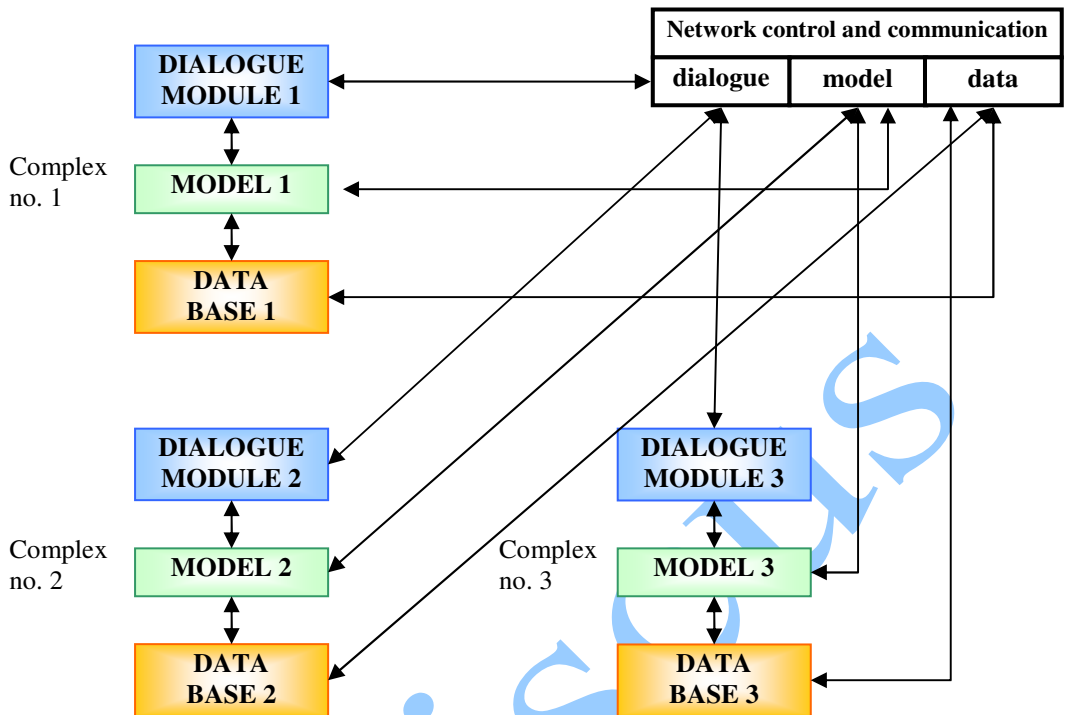

**Figure 4.** *The network architecture*

**The centered architecture**

In case of the centered architecture, each of the models (Figure 5.) depends on a single dialogue module and communicates with a single data base. The existence of a unique dialogue module is however an advantage for the user and the unique data base is considerably improving the information exchange between the models.

This architecture's lacks in flexibility in case a different new model would be introduced.

**Hierarchy architecture**

Unlike the centered architecture, in the hierarchy architecture situation (Figure 6), the dialogue module is divided, and to the data base module a supplementary level is added, in order to increase the system's adaptability by facilitating the adding of new models.

344

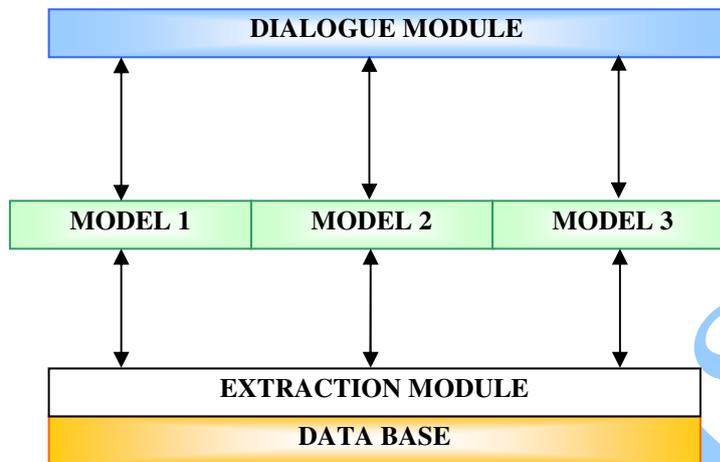

**Figure 5.** *Centered architecture*

The dialogue module is formed by dialogue modules directly connected to certain models, and the supervisor module, that is interposing between the user's dialogue modules and the models, managing the information exchange between the dialogue modules and the models.

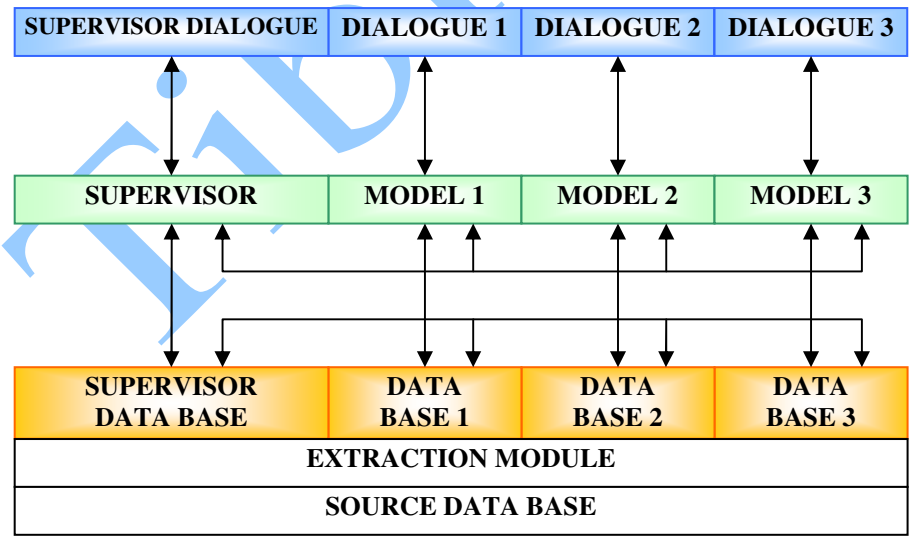

**Figure 6.** *Hierarchy architecture*

The n hierarchy architecture is very often used considering that the models can have different structures. The user's control is easily done with the supervisor that has its own dialogue resources, so the user would have fully control when passing from a model to another. The data exchange between models is also solved by the supervisor that reads data from the models, copies them into the data base, and then is transferring them to other models.

The network architecture, the centered architecture and the hierarchy architecture have each proved advantages and disadvantages, which are shown in Table 1.

**Table 1.**

Advantage and disadvantage of decision support systems architectures

| ARCHITECTURE TYPE | ADVANTAGE | DISADVANTAGE |
| --- | --- | --- |
| **NETWORK** | - open architecture<br>- high modularity | - weak integration<br>- lack of dialogue unit<br>- difficult data exchange<br>- difficult control design |
| **CENTERED** | - high integration<br>- dialogue unity<br>- facilitated data exchange<br>- easy to make | - difficult to modify<br>- difficult new model adding<br>- low confidentiality data access |
| **HIERARCHY** | - high integration<br>- dialogue unity<br>- facilitated data base development<br>- facilitated changes<br>- user friendly | - difficult to realize the supervisor and the extraction module |

**Conclusions**

As a conclusion, the hierarchy architecture seems to include the advantages of the other two architectures. The dialogue units, the facilitated data exchange, the optimum integration from the centered architecture are also advantages of the hierarchy architecture. Thanks to the data extraction module and the supervisor, one can easily add new models.